\documentclass[pra,aps,twocolumn,superscriptaddress,amsmath,amssymb]{revtex4-1}

\usepackage{amsthm}
\newtheorem{theorem}{Theorem}

\begin{document}

\title[The quantum separability problem]{The quantum separability problem is a simultaneous hollowisation matrix analysis problem}

\author{A. Neven}
\affiliation{Institut de Physique Nucl\'eaire, Atomique et de Spectroscopie, \\ CESAM, University of Liege, B\^at.~B15, Sart Tilman, Li\`ege 4000, Belgium}

\author{T. Bastin}
\affiliation{Institut de Physique Nucl\'eaire, Atomique et de Spectroscopie, \\ CESAM, University of Liege, B\^at.~B15, Sart Tilman, Li\`ege 4000, Belgium}

\date{28 June 2018}

\begin{abstract}
We use the generalized concurrence approach to investigate the general multipartite separability problem. By extending the preconcurrence matrix formalism to arbitrary multipartite systems, we show that the separability problem can be formulated equivalently as a pure matrix analysis problem that consists in determining whether a given set of symmetric matrices is simultaneously unitarily congruent to hollow matrices, i.e., to matrices whose main diagonal is only composed of zeroes.
\end{abstract}

\maketitle

\section{Introduction}

Quantum entanglement is at the heart of quantum mechanics and intimately linked to its nonlocal feature~\cite{HorodeckiReview}. It is a key resource in many promising applications, like, to cite a few, quantum cryptography~\cite{Eck91}, quantum communication~\cite{Zei07}, quantum imaging~\cite{Tei01}, or also quantum sensing~\cite{Wein10}. In this context, the ability to distinguish both experimentally and theoretically between entangled and separable states is a crucial issue. Theoretically, this issue is entirely solved in the pure state case where general and practical necessary and sufficient separability criteria have been identified (see, e.g., Ref.~\cite{Yu2006}). As a reminder, a pure state is said separable if it can be written as a tensor product of individual party states and is entangled otherwise. For mixed states, the separability question is much more involved and remains open in the very general case (a mixed state is separable if it can be written as a convex sum of projectors onto separable pure states and entangled otherwise). Still various necessary but not sufficient conditions of separability have been stated~\cite{HorodeckiReview, GuhneToth}, such as the positive partial transpose (PPT) criterion~\cite{Peres}, combinatorially independent permutation criteria~\cite{HorodeckiRealignment,Wocjan}, Bell-type inequalities \cite{Seevinck}, or criteria based on entanglement witnesses~\cite{HorodeckiPPT,Terhal}. In some restricted cases, some of these above-cited criteria turn out to be also sufficient conditions of separability. This happens for example for the PPT criterion in low-dimensional or low-rank cases, such as for qubit-qubit or qubit-qutrit systems~\cite{HorodeckiPPT}, for $\mathbb{C}^m \otimes \mathbb{C}^n (m\leq n)$ bipartite states with rank at most $n$ \cite{HorodeckiLewenstein}, or even for general multipartite mixed states with rank at most 3 \cite{Chen2013}.

The concurrence~\cite{Wootters} is another tool that proved to provide a necessary and sufficient condition of separability in 2-qubit systems. It is defined for pure states $|\psi\rangle$ as
\begin{equation}
C(\psi) = |\langle \psi | S |\psi^* \rangle |, \label{concurrence}
\end{equation}
where $S = \sigma_y \otimes \sigma_y$ is the 2-qubit spin-flip operator with $\sigma_y$ the second Pauli matrix and where $|\psi^*\rangle$ is the complex conjugate of $|\psi\rangle$ expressed in the computational basis. For mixed states $\rho$, the concurrence is defined via the standard convex-roof construction~:
\begin{equation}
C(\rho)=\inf_{ \{p_k,|\psi_k\rangle \} } \sum_k{p_k C(\psi_k)}, \label{convex-roof}
\end{equation}
where the infimum is computed over all possible decompositions of $\rho$, i.e., all sets $\{p_k,|\psi_k\rangle \}$ such that $\rho=\sum_k p_k |\psi_k \rangle \langle \psi_k|$. The concurrence is an entanglement measure that vanishes only for separable states~\cite{Wootters} and this provides an easy necessary and sufficient condition of separability~: a state $\rho$ is separable if and only if $C(\rho)=0$. In general, the minimization implied by convex-roofs is a very challenging task. However, in the case of the concurrence, Eq.~(\ref{convex-roof}) simplifies to~\cite{Wootters}
\begin{equation}
C(\rho) = \max\{0,\lambda_1 - \lambda_2 - \lambda_3 - \lambda_4\},
\label{formulaWootters}
\end{equation}
with $\lambda_i \: (i=1,\dots,4)$ the square roots of the eigenvalues of $\rho S  \rho^* S $ sorted in decreasing order.

The 2-qubit concurrence has been generalized to more general bipartite~\cite{Audenaert} or even multipartite~\cite{Yu2006} systems by the introduction of a set of generalized concurrences $C_\alpha$ ($\alpha = 1, 2, \ldots$) defined similarly as in Eq.~(\ref{concurrence}) but each with a specific generalized ``spin-flip" operator $S_\alpha$~\cite{Yu2006, Audenaert}. The cancellation of all generalized concurrences still provides a necessary and sufficient separability condition, however only for pure states. Though the extension to mixed states via the convex-roof construction yields a similar elegant result as in Eq.~\eqref{formulaWootters} for each $C_\alpha$~\cite{Yu2006, Audenaert}, the cancellation of all of them only provides a necessary separability condition for mixed states~\cite{Yu2006, Audenaert}. In this paper, we show that the missing element to get a necessary and sufficient condition of separability based on generalized concurrences can be formalized equivalently as a pure matrix analysis problem that consists in \emph{determining whether a given set of symmetric matrices is simultaneously unitarily congruent to hollow matrices, i.e., to matrices whose main diagonal is composed only of zeroes}.

To this aim, we first refine in Sec.~\ref{pureStateSection} the necessary and sufficient condition (NSC) of separability based on generalized concurrences for pure state by showing how to get an optimal non-redundant set of generalized ``spin-flip" operators $S_{\alpha}$ for arbitrary multipartite systems. We then extend in Sec.~\ref{mixedStateSection} the concept of preconcurrence matrices~\cite{Badziag} to these operators and address the mixed state case. We finally draw conclusions in Sec.~\ref{Concl}.

\section{Pure-state case}
\label{pureStateSection}

The generalized ``spin-flip" operators $S_{\alpha}$ introduced in Refs.~\cite{Yu2006, Audenaert} are generated either from tensor products of SO($n$) generators~\cite{Yu2006} or from $2 \times 2$ minor equations from tensor matricizations~\cite{Audenaert, Gillet}. Both methods unfortunately produce highly redundant sets of operators. Here, we show how to extract from them the only independent ones. For this purpose we make use of the $2 \times 2$ minor equation method~\cite{Audenaert} that is better suited for this task. We consider an arbitrary multipartite system with Hilbert space $\mathcal{H}=\mathcal{H}_1 \otimes \mathcal{H}_2 \otimes \cdots \otimes \mathcal{H}_N$, where $\mathcal{H}_j$ ($j = 1, \ldots,N$) are the individual Hilbert spaces of dimension $m_j \geq 2$ for each party. Each $\mathcal{H}_j$ is isomorphic to $\mathbb{C}^{m_j}$. In a computational basis $|\mathbf{i}\rangle \equiv |i_1, \dots, i_N\rangle \equiv |i_1\rangle \otimes \cdots \otimes |i_N\rangle$, with $i_j = 0, \ldots, m_j-1$ ($j = 1, \ldots, N$), any pure state $|\psi\rangle$ can be expressed as
\begin{equation}
|\psi\rangle = \sum_{\mathbf{i}} a_{\mathbf{i}} |\mathbf{i}\rangle \equiv \sum_{i_1=0}^{m_1-1} \cdots \sum_{i_N=0}^{m_N-1} a_{i_1,\dots,i_N} \; |i_1, \dots, i_N\rangle.
\end{equation}
The state $|\psi\rangle$ is separable if and only if the $N$-order tensor $A$ with components $a_{\mathbf{i}} \equiv a_{i_1,\dots,i_N}$ is of rank 1~\cite{tensor}. This is the case if and only if all mode-$k$ matricizations $A^{(k)}$ of $A$ ($k=1,\dots,N$) are themselves of rank 1~\cite{tensor}. The mode-$k$ matricization $A^{(k)}$ of $A$ is the $m_k \times \mathrm{dim} \mathcal{H}/m_k$ matrix whose columns are indexed by all possible values of $\mathbf{i}_{\neg k} \equiv (i_1, \ldots, i_{k-1}, i_{k+1}, \ldots, i_N)$ and filled with the corresponding elements $a_{\mathbf{i}}$ with $i_k$ ranging from $0$ to $m_k-1$. The $N$ matricizations of $A$ are of rank 1 if and only if all their $2 \times 2$ minors vanish, i.e., if and only if
\begin{equation}
a_{\mathbf{i}} a_{\mathbf{i}^\prime} = a_{\mathbf{i}_{[i_k^\prime]}} a_{\mathbf{i}_{[i_k]}^\prime}, \quad \forall k, \forall \mathbf{i}, \mathbf{i}^\prime : i^\prime_k > i_k, \mathbf{i}_{\neg k}^\prime > \mathbf{i}_{\neg k}, \label{eq_aa=aa}
\end{equation}
where $\mathbf{i}_{[i_k^\prime]} \equiv (i_1, \ldots, i_{k-1}, i_k^\prime, i_{k+1}, \ldots, i_N)$, $\mathbf{i}_{[i_k]}^\prime \equiv (i_1^\prime, \ldots, i_{k-1}^\prime, i_k, i_{k+1}^\prime, \ldots, i_N^\prime)$, and $\mathbf{i}_{\neg k}^\prime > \mathbf{i}_{\neg k}$ means that at least one component of $\mathbf{i}_{\neg k}^\prime$ differs from its equivalent in $\mathbf{i}_{\neg k}$ and that the first of these differing components is greater for $\mathbf{i}_{\neg k}^\prime$ than for $\mathbf{i}_{\neg k}$. If we introduce the generalized concurrences
\begin{equation}
C_{k,\mathbf{i},\mathbf{i}^\prime}(\psi) \equiv |\langle \psi | S_{k, \mathbf{i}, \mathbf{i}^\prime} | \psi^* \rangle |
\end{equation}
with
\begin{equation}
S_{k, \mathbf{i}, \mathbf{i}^\prime} = |\mathbf{i}\rangle\langle\mathbf{i}^\prime| - |\mathbf{i}_{[i_k^\prime]}\rangle\langle\mathbf{i}_{[i_k]}^\prime| + \mathrm{h.c.},
\end{equation}
Eq.~(\ref{eq_aa=aa}) is equivalent to
\begin{equation}
    C_{k,\mathbf{i},\mathbf{i}^\prime}(\psi) = 0, \quad \forall k, \forall \mathbf{i}, \mathbf{i}^\prime : i^\prime_k > i_k, \mathbf{i}_{\neg k}^\prime > \mathbf{i}_{\neg k}. \label{Cvanish}
\end{equation}
This expresses an NSC of separability for the multipartite pure state $|\psi\rangle$.

The number of generalized concurrences implied by Eq.~(\ref{Cvanish}) amounts to $\sum_{k=1}^N {m_k \choose 2} {\mathrm{dim} \mathcal{H}/m_k \choose 2}$, that simplifies to $Nd^N(d^{N-1} - 1)(d-1)/4$ for an $N$-qudit system ($m_k = d$, $\forall k$). Actually, many of these concurrences are redundant if not identical and they do not cancel independently of each other. It is useful to identify a minimal set of these equalities that provides equivalently an NSC of separability. To this aim, we first introduce some notations. In Eq.~\eqref{eq_aa=aa}, the conditions $i^\prime_k > i_k$ and $\mathbf{i}_{\neg k}^\prime > \mathbf{i}_{\neg k}$ imply that the indexes $\mathbf{i}$ and $\mathbf{i}^\prime$ in an equality necessarily differ in a number of components greater than or equal to 2. Let $q_{\mathbf{i}, \mathbf{i}^\prime}$ be this number of different components and $\mathcal{Q}_{\mathbf{i},\mathbf{i}^\prime}$ the set gathering their positions~:  $\mathcal{Q}_{\mathbf{i},\mathbf{i}^\prime} = \{ \: k : \: i_k \neq i_k^\prime \}$ and $q_{\mathbf{i}, \mathbf{i}^\prime} = \#\mathcal{Q}_{\mathbf{i},\mathbf{i}^\prime}$. The (possibly empty) complement $\overline{\mathcal{Q}}_{\mathbf{i},\mathbf{i}^\prime}$ is the set $\{ \: k \: : \: i_k=i_k^\prime \}$. We then define the two $q_{\mathbf{i}, \mathbf{i}^\prime}$-tuples $ \mathbf{d}_{\mathbf{i},\mathbf{i}^\prime} \equiv (i_{(\mathcal{Q}_{\mathbf{i},\mathbf{i}^\prime})_1}, \dots, i_{(\mathcal{Q}_{\mathbf{i},\mathbf{i}^\prime})_{q_{\mathbf{i}, \mathbf{i}^\prime}}})$ and $ \mathbf{d}_{\mathbf{i},\mathbf{i}^\prime}^\prime \equiv (i^\prime_{(\mathcal{Q}_{\mathbf{i},\mathbf{i}^\prime})_1}, \dots, i^\prime_{(\mathcal{Q}_{\mathbf{i},\mathbf{i}^\prime})_{q_{\mathbf{i}, \mathbf{i}^\prime}}})$, as well as, if $q_{\mathbf{i}, \mathbf{i}^\prime} \neq N$, the $(N-q_{\mathbf{i}, \mathbf{i}^\prime})$-tuple $\mathbf{c}_{\mathbf{i},\mathbf{i}^\prime} \equiv (i_{(\overline{\mathcal{Q}}_{\mathbf{i},\mathbf{i}^\prime})_1}, \dots, i_{(\overline{\mathcal{Q}}_{\mathbf{i},\mathbf{i}^\prime})_{N-q_{\mathbf{i}, \mathbf{i}^\prime}}})$, where $(\mathcal{A})_k$ ($\mathcal{A} = \mathcal{Q}_{\mathbf{i},\mathbf{i}^\prime}, \overline{\mathcal{Q}}_{\mathbf{i},\mathbf{i}^\prime}$) denotes the $k$-th element of the set $\mathcal{A}$.

We then structure the set of equalities of Eq.~(\ref{eq_aa=aa}) into subsets $\mathcal{S}_{\mathcal{Q}, \mathbf{c}, (\mathbf{d},\mathbf{d}^\prime)}$ that each gather all equalities with index couples ($\mathbf{i}$,$\mathbf{i}^\prime$) said $\mathcal{S}_{\mathcal{Q}, \mathbf{c}, (\mathbf{d},\mathbf{d}^\prime)}$-compatible, i.e., such that $\mathcal{Q}_{\mathbf{i}, \mathbf{i}^\prime} = \mathcal{Q}$, $\mathbf{c}_{\mathbf{i}, \mathbf{i}^\prime} = \mathbf{c}$ and $(\mathbf{d}_{\mathbf{i}, \mathbf{i}^\prime}, \mathbf{d}^\prime_{\mathbf{i}, \mathbf{i}^\prime}) = (\mathbf{d},\mathbf{d}^\prime)$ up to swaps of $\mathbf{d}^\prime$ components with their related components in $\mathbf{d}$. We have $\#\mathcal{S}_{\mathcal{Q}, \mathbf{c}, (\mathbf{d},\mathbf{d}^\prime)} = q 2^{q-2}$, with $q = \#\mathcal{Q}$. In each subset $\mathcal{S}$, one easily checks that
the number of distinct pairs $\{\mathbf{i}, \mathbf{i}^\prime\}$ and $\{\mathbf{i}_{[i'_k]}, \mathbf{i}^\prime_{[i_k]}\}$ amounts together to $2^{q - 1}$ and that the number of independent equalities is equal to $2^{q - 1} - 1$. The independent equalities of each subset $\mathcal{S}$ keep all independent between each other when the subsets are grouped together. To see this, let us consider an arbitrary equality indexed by ($k,\mathbf{i}, \mathbf{i}^\prime$) in a subset $\mathcal{S}$ and let us show that it is independent of all equalities of any other subsets $\mathcal{S}^\prime$. We consider the state $|\psi\rangle = a_\mathbf{i} |\mathbf{i}\rangle + a_{\mathbf{i}^\prime} |\mathbf{i}^\prime\rangle$. For this state, all equalities of any subsets $\mathcal{S}^\prime \neq \mathcal{S}$ are trivially satisfied since they read $0 = 0$, while the equality ($k,\mathbf{i}, \mathbf{i}^\prime$) of $\mathcal{S}$ reads $a_\mathbf{i} a_{\mathbf{i}^\prime} = 0$. Hence, the separability of the state (which requires here $a_\mathbf{i} = 0$ or $a_{\mathbf{i}^\prime} = 0$) can only be certified with help of this latter equality that cannot therefore be skipped. It follows that the total number of independent equalities amounts to
\begin{equation}
Q = \sum_{q=2}^N {\left( 2^{q-1}-1 \right) \sum_{\renewcommand{\arraystretch}{0.7} \begin{array}{c} \scriptstyle \mathcal{Q}_{\mathbf{i},\mathbf{i}^\prime} : \\ \scriptstyle q_{\mathbf{i},\mathbf{i}^\prime} = q \end{array} \renewcommand{\arraystretch}{1}} {\prod_{k=1}^q {m_{(\mathcal{Q}_{\mathbf{i},\mathbf{i}^\prime})_k} \choose 2} \prod_{k=1}^{N-q} m_{(\overline{\mathcal{Q}}_{\mathbf{i},\mathbf{i}^\prime})_k}}},
\end{equation}
that simplifies to
\begin{equation}
    Q = d^{N+1}\frac{d-1}{4}\left(1 - 2\left(1 + \frac{1}{d}\right)^N + \left(1 + \frac{2}{d}\right)^N\right)
\end{equation}
for an $N$-qudit system.

A set of equivalent independent equalities is obtained if in each subset $\mathcal{S}$ we rather consider the equalities $a_\mathbf{i} a_{\mathbf{i}^\prime} = a_\mathbf{j} a_{\mathbf{j}^\prime}$, with ($\mathbf{i}$,$\mathbf{i}^\prime)$ any fixed $\mathcal{S}$-compatible index couple and $(\mathbf{j},\mathbf{j}^\prime)$ all possible $\mathcal{S}$-compatible index couples distinct from $(\mathbf{i},\mathbf{i}^\prime)$ and such that $(\mathbf{d}_{\mathbf{j}, \mathbf{j}^\prime}, \mathbf{d}^\prime_{\mathbf{j}, \mathbf{j}^\prime}) = (\mathbf{d}_{\mathbf{i}, \mathbf{i}^\prime}, \mathbf{d}^\prime_{\mathbf{i}, \mathbf{i}^\prime})$ up to swaps of any components but the last of $\mathbf{d}^\prime_{\mathbf{i}, \mathbf{i}^\prime}$ with their equivalents in $\mathbf{d}_{\mathbf{i}, \mathbf{i}^\prime}$. All these equalities can be equivalently written
\begin{equation}
C_{\mathbf{i},\mathbf{i}^\prime,\mathbf{j},\mathbf{j}^\prime}(\psi) = 0,
\label{Calphavanish}
\end{equation}
with the generalized concurrences
\begin{equation}
C_{\mathbf{i},\mathbf{i}^\prime,\mathbf{j},\mathbf{j}^\prime}(\psi) = |\langle \psi | S_{\mathbf{i},\mathbf{i}^\prime,\mathbf{j},\mathbf{j}^\prime} | \psi^* \rangle|,
\label{Ciipjjp}
\end{equation}
where
\begin{equation}
S_{\mathbf{i},\mathbf{i}^\prime,\mathbf{j},\mathbf{j}^\prime} = |\mathbf{i}\rangle\langle\mathbf{i}^\prime| - |\mathbf{j}\rangle\langle\mathbf{j}^\prime| + \mathrm{h.c.}
\label{Siipjjp}
\end{equation}
The equalities (\ref{Calphavanish}) are hereafter merely indexed by the subscript $\alpha = 1, \ldots, Q$ and similarly for all related generalized concurrences~(\ref{Ciipjjp}) and hermitian operators (\ref{Siipjjp}). The NSC of separability (\ref{Cvanish}) for pure states can then be refined accordingly~:
\begin{theorem}
A general $N$-partite pure state $|\psi\rangle$ in $\mathcal{H} = \mathbb{C}^{m_1} \otimes \cdots \otimes \mathbb{C}^{m_N}$ ($m_j \geq 2$) is separable if and only if $C_\alpha(\psi) = 0$, $\forall \alpha=1,\dots,Q$.
\end{theorem}

Even though the number $Q$ of independent generalized concurrences $C_\alpha$ quickly grows with the number of parties and dimensions of the subsystem spaces, the elimination of all redundancies is not anecdotic. For $N$ qubits, the numbers of generalized concurrences implied in the separability criterion of Eq.~(\ref{Cvanish}) amount to $2$, $18$, $112$, $600$, $2976$, $14112$, $65024$, $293760$, and $1308160$ for $N = 2$ to $10$, respectively. In contrast, the numbers of true independent concurrences among them amount to $Q = 1$, $9$, $55$, $285$, $1351$, $6069$, $26335$, $111645$, and $465751$, respectively. For 2 qubits, the formalism defines the single independent concurrence $C_1 = |\langle \psi | S_1 | \psi^*\rangle|$ with $S_1 = |00\rangle\langle11| - |10\rangle\langle01| + \mathrm{h.c.} = \sigma_y \otimes \sigma_y$. This is nothing but the Wootters' concurrence (\ref{concurrence})~\cite{Wootters}.

\section{Mixed-state case}
\label{mixedStateSection}

We now turn to mixed states. The generalized concurrences $C_\alpha$ are extended to mixed states using the standard convex-roof construction. For all $\alpha$, we define
\begin{equation}
C_\alpha(\rho) = \inf_{ \{p_k,|\psi_k\rangle \} } \sum_k{p_k \: C_\alpha(\psi_k)},
\end{equation}
where the infimum is taken over all possible decompositions of $\rho$. If a state $\rho$ is separable, a decomposition exists where each state $|\psi_k\rangle$ of the decomposition is separable. This implies $C_\alpha(\psi_k) = 0, \forall \alpha, k$ and hence $C_\alpha(\rho) = 0, \forall \alpha$. The converse is not true. The cancellation of all concurrences $C_\alpha(\rho)$ implies that for each of them a decomposition exists where the concurrence in question vanishes for each state of the decomposition. For each concurrence however, the decomposition in question may vary. Therefore, the cancellation of all concurrences $C_\alpha(\rho)$ does not imply that a decomposition exists where all concurrences of each state of the decomposition would vanish, in which case all states of the decomposition would be separable and hence the mixed state itself. The cancellation of all concurrences is a necessary but not sufficient condition of separability for mixed states~:
\begin{equation}
    \rho \,\, \mathrm{separable} \Rightarrow C_\alpha(\rho) = 0, \forall \alpha.
\end{equation}
For instance, the 3-qubit mixed state $\rho = (|D_3^{(0,2)}\rangle\langle D_3^{(0,2)}| + |D_3^{(1,2)}\rangle\langle D_3^{(1,2)}| + |D_3^{(1,3)}\rangle\langle D_3^{(1,3)}|)/3$, with $|D_3^{(k,k')}\rangle \equiv (|D_3^{(k)}\rangle + |D_3^{(k')}\rangle)/\sqrt{2}$ where $|D_3^{(k)}\rangle$ ($k = 0, \ldots, 3$) denote the 3-qubit Dicke states~\cite{footnote}, is a negative partial transpose (NPT) state and hence entangled~\cite{Peres} although all concurrences $C_{\alpha}(\rho)$ ($\alpha = 1, \ldots, 9$) vanish for this state (these concurrences are easily computed using Eq.~(\ref{Calpha}) hereafter).

To obtain a necessary and sufficient separability condition, a deeper analysis of the possible decompositions of the mixed states is required. We first provide an easy necessary and sufficient condition for the cancellation of a given individual concurrence $C_{\alpha}(\rho)$. To this aim, we generalize to arbitrary multipartite systems the preconcurrence matrix formalism introduced in Refs.~\cite{Wootters, Audenaert, Badziag} for bipartite systems. Let $\mathcal{D} = \{p_k, |\psi_k\rangle, k = 1, \ldots, p \}$ be a decomposition of the mixed state $\rho$~: $\rho = \sum_{k=1}^p p_k |\psi_k\rangle \langle \psi_k|$. We introduce the unnormalized states $|\tilde{\psi}_k\rangle \equiv \sqrt{p_k} |\psi_k\rangle$ so as to write $\rho = \sum_{k=1}^p |\tilde{\psi}_k\rangle \langle \tilde{\psi}_k|$. The \emph{preconcurrence matrix} $\tau_{\alpha}^\mathcal{D}$ is defined as the square $p \times p$ matrix of elements
\begin{equation}
    (\tau_{\alpha}^\mathcal{D})_{i j} = \langle \tilde{\psi}_i | S_{\alpha} | \tilde{\psi}_j^* \rangle, \quad i,j = 1, \ldots, p.
\end{equation}
The preconcurrence matrix is symmetric~: $(\tau_{\alpha}^\mathcal{D})^T = \tau_{\alpha}^\mathcal{D}$. Its diagonal elements are the concurrences (up to the absolute value) of the (unnormalized) states $|\tilde{\psi}_k\rangle$ of the decomposition, hence the name of the matrix.

Of particular interest is the eigendecomposition $\mathcal{E}$ of $\rho$, i.e., the decomposition of $\rho$ over its eigenvectors $|v_k\rangle$ with nonzero eigenvalues $\lambda_k$~: $\rho = \sum_{k = 1}^r \lambda_k |v_k\rangle\langle v_k| = \sum_{k = 1}^r |\tilde{v}_k\rangle\langle\tilde{v}_k|$, with $|\tilde{v}_k\rangle = \sqrt{\lambda_k} |v_k\rangle$. Here $r$ is the rank of $\rho$. We set $\tau_{\alpha} \equiv \tau_{\alpha}^\mathcal{E}$. No other decomposition of $\rho$ can contain a number of states smaller than $r$~\cite{Hughston}. For an arbitrary decomposition $\mathcal{D} = \{p_k, |\psi_k\rangle, k = 1, \ldots, p \}$ of $\rho$, a $p \times p$ unitary matrix $U$ always exists such that $(|\tilde{\psi}_1\rangle, \ldots, |\tilde{\psi}_p\rangle)^T = U^* (|\tilde{v}_1\rangle, \ldots, |\tilde{v}_p\rangle)^T$ with $|\tilde{v}_k\rangle \equiv 0$ for $k > r$~\cite{Hughston}. Hence the matrix $\tau_{\alpha}^{\mathcal{D}}$ is unitarily congruent to the matrix $\tau_{\alpha}$ extended with $p - r$ rows and columns only composed of zeroes~: $\tau_\alpha^{\mathcal{D}} = U \tau_{\alpha}^{\mathrm{ext}} U^T$, with $\tau_{\alpha}^{\mathrm{ext}}$ the so extended $\tau_{\alpha}$ matrix. Conversely, any $p \times p$ unitary $U^*$ applied on $(|\tilde{v}_1\rangle, \ldots, |\tilde{v}_p\rangle)^T$ defines an alternative decomposition of $\rho$ (over a number of vectors comprised between $r$ and $p$)~\cite{Hughston} and the preconcurrence matrix related to this decomposition is directly obtained by the corresponding unitary congruence $U \tau_{\alpha}^{\mathrm{ext}} U^T$.

We can now provide a necessary and sufficient condition to have $C_\alpha(\rho) = 0$ for a given $\alpha$. The concurrence $C_\alpha(\rho)$ vanishes if and only if a decomposition $\mathcal{D}$ of $\rho$ exists where the concurrence of each state of the decomposition vanishes, or said differently, where the preconcurrence matrix $\tau_\alpha^{\mathcal{D}}$ is hollow (its diagonal is only composed of zeroes). This is the case if and only if there exists an extension $\tau_{\alpha}^{\mathrm{ext}}$ of $\tau_{\alpha}$ that is unitarily congruent to a hollow matrix, i.e., that has singular values $s_k$ that, when sorted in decreasing order, verify $s_1 - \sum_{k=2}^p s_k \leq 0$ (Thompson's Theorem~1 and Lemma~2~\cite{Thompson} on the conditions of existence of symmetric matrices with prescribed singular values and diagonal elements). Since the nonzero singular values of any extension $\tau_{\alpha}^{\mathrm{ext}}$ of $\tau_\alpha$ are exactly the same as those of $\tau_\alpha$ itself, we get merely that $C_\alpha(\rho) = 0$ if and only if the singular values $s_1,\dots,s_r$ of $\tau_\alpha$, sorted in decreasing order, verify $s_1 - \sum_{k=2}^r s_k \leq 0$. In this case the symmetric matrix $\tau_\alpha$ (or its $4 \times 4$ extension in the special case (SC) $r = 3$ and $s_1 - s_2 - s_3 < 0$) is itself unitarily congruent to a hollow matrix~\cite{Thompson} (we say \emph{hollowisable} by unitary congruence) and a decomposition of $\rho$ over exactly $r$ (or $r+1$ in the SC) vectors $|\psi_k\rangle$ with $C_\alpha(\psi_k) = 0, \forall k$ is ensured to exist. In the SC, the symmetric matrix $\tau_\alpha$ is itself not hollowisable by unitary congruence and a decomposition of $\rho$ over exactly $r = 3$ vectors $|\psi_k\rangle$ with $C_\alpha(\psi_k) = 0, \forall k$ does not exist. Since the nonzero singular values of $\tau_\alpha$ are identical to the square roots of the nonzero eigenvalues of $\rho S_\alpha \rho^* S_\alpha$, all this implies the following Theorem~:
\begin{theorem}
A given concurrence $C_\alpha$ of a general $N$-partite mixed state $\rho$ vanishes if and only if the (symmetric) $r \times r$ preconcurrence matrix $\tau_\alpha \equiv \tau_\alpha^\mathcal{E}$ (or its $4 \times 4$ extension if $r = 3$) is hollowisable by unitary congruence ($r = \mathrm{rank}\rho$). This is the case if and only if the singular values $s_1,\dots,s_r$ of $\tau_\alpha$ (or equivalently the square roots of the $r$ largest eigenvalues of $\rho S_\alpha \rho^* S_\alpha$), sorted in decreasing order, satisfy
\begin{equation}
\label{HollowisabilityConditionEq}
 s_1 - \sum_{k=2}^r{s_k} \leq 0.
\end{equation}
\end{theorem}

The singular values $s_1,\dots,s_r$ of $\tau_\alpha$ (or equivalently the square roots of the $r$ largest eigenvalues of $\rho S_\alpha \rho^* S_\alpha$), sorted in decreasing order, actually fully characterize the generalize concurrence $C_\alpha(\rho)$. We have
\begin{equation}
C_\alpha(\rho) = \max(0,s_1 - \sum_{k=2}^r{s_k}).
\label{Calpha}
\end{equation}
Indeed, we can first write $C_\alpha(\rho) = \min_{\mathcal{D} = \{p_k,|\psi_k\rangle\}} \sum_{k=1}^p p_k C_\alpha(\psi_k) = \min_{\mathcal{D}} \sum_k C_\alpha(\tilde{\psi}_k) = \min_{\mathcal{D}} \sum_k |(\tau_\alpha^\mathcal{D})_{kk}| = \min_{p \geq r, U \in \mathrm{U}(p)} \sum_k |(U\tau_\alpha^\mathrm{ext}U^T)_{kk}|$, with $\mathrm{U}(p)$ the $p \times p$ unitary group. According to Thompson's Theorem~1 and Lemma~2~\cite{Thompson} (hereafter referred to as Thompson's Theorem), the diagonal elements $d_k$ ($k = 1, \ldots, p$) of any unitarily congruent matrix of $\tau_\alpha^{\mathrm{ext}}$ verify, if sorted in decreasing order of their absolute values, $\sum_{k=1}^{i-1} |d_k| - \sum_{k=i}^p |d_k| \leq \sum_{k=1, k\neq i}^p s_k - s_i$ for $1 \leq i \leq p$, where $s_1, \ldots, s_p$ are the singular values of $\tau_\alpha^{\mathrm{ext}}$ sorted in decreasing order. For $i = 1$, this condition implies that the sum $\sum_{k=1}^p |d_k|$ is lower bounded by $s_1 - \sum_{k=2}^p {s_k}$. This sum is also trivially lower bounded by 0, and hence by $\max(0,s_1 - \sum_{k=2}^p{s_k})$. This lower bound is realized. To see this, it is enough to observe that Thompson's Theorem~\cite{Thompson} allows for the existence of a unitarily congruent matrix to $\tau_\alpha^{\mathrm{ext}}$ with diagonal elements $d_1 = \max(0,s_1 - \sum_{k=2}^p s_k)$ and $d_2 = \cdots = d_p = 0$. Since the nonzero singular values of any extension $\tau_\alpha^{\mathrm{ext}}$ of $\tau_\alpha$ are the same as those of $\tau_\alpha$ itself, the conclusion follows immediately.

We finally address the separability question of a mixed state $\rho$. This separability implies the existence of a decomposition $\mathcal{D}$ of $\rho$ over separable states $|\psi_k\rangle$, whose number, $p$, is  necessarily comprised between $r$ and $r^2$~\cite{Uhlmann}. In this case, if $U^*$ is the unitary that transforms the list of unnormalized eigenvectors of $\rho$ (extended with $p - r$ null vectors) to the unnormalized separable state list, the preconcurrence matrices $\tau_\alpha^\mathcal{D}$ are hollow, $\forall \alpha$, and unitarily congruent with the unitary $U$ to the $p \times p$ extensions $\tau_\alpha^{\mathrm{ext}}$ of $\tau_\alpha$, respectively. In other words, there exists a number $p$ between $r$ and $r^2$ such that the $p \times p$ extensions $\tau_\alpha^{\mathrm{ext}}$ of $\tau_\alpha$ are all hollowisable by unitary congruence with the \emph{same} unitary $U$. The converse is true. If this number $p$ exists, a decomposition of $\rho$ over $p$ states with vanishing concurrences for all $\alpha$ exists, hence over $p$ separable states. We thus have the following necessary and sufficient condition of separability for mixed states~:
\begin{theorem}
\label{genTh}
A general $N$-partite mixed state $\rho$ is separable if and only if a number $p$ comprised between $r$ and $r^2$ can be found so that the $p \times p$ extensions $\tau_\alpha^{\mathrm{ext}}$, $\forall \alpha$, are all \emph{simultaneously} hollowisable by unitary congruence, i.e., with the \emph{same} unitary $U$ ($r = \mathrm{rank} \rho$). In this case, a separable decomposition of $\rho$ is given by $\rho = \sum_{k=1}^p |\tilde{\psi}_k\rangle\langle \tilde{\psi}_k|$, with $(|\tilde{\psi}_1\rangle, \ldots, |\tilde{\psi}_p\rangle)^T = U^* (|\tilde{v}_1\rangle, \ldots, |\tilde{v}_p\rangle)^T$.
\end{theorem}

This Theorem shows that the general separability problem of mixed states is equivalent to a pure matrix analysis problem that consists in determining whether a given set of symmetric matrices determined by the mixed states is simultaneously unitarily congruent to hollow matrices. In the same way that quantum compatibility of observables is equivalent to simultaneous diagonalisability of hermitian matrices, quantum separability is equivalent to simultaneous hollowisability of symmetric matrices. While hollowisability by unitary congruence of a single symmetric matrix is a well-known problem that is addressed via the inequality (\ref{HollowisabilityConditionEq}), the simultaneous hollowisability by unitary congruence of several symmetric matrices remains an open matrix analysis problem (recent results about the simultaneous unitary congruence problem can be found in Refs.~\cite{AlpinIkramov,Gerasimova}). For $2 \times 2$ matrices, one easily shows from the general form of $2 \times 2$ unitaries that a set of symmetric matrices are simultaneously hollowisable by unitary congruence if and only if they are all individually hollowisable by unitary congruence (which happens iff their two singular values are identical) and proportional to each other. The same holds for $2 \times 2$ matrices that are extended with an identical number of rows and columns only composed of zeroes~: they are simultaneously hollowisable by unitary congruence if and only if they are all individually hollowisable by unitary congruence (which happens here iff their two largest singular values are identical) and proportional to each other. As a consequence, for rank-2 mixed states, the conditions of simultaneous hollowisability by unitary congruence of all preconcurrence matrices $\tau_\alpha$, $\forall \alpha$, or of all $3 \times 3$ or $4 \times 4$ extensions $\tau_\alpha^{\mathrm{ext}}$ of $\tau_\alpha$ are equivalently satisfied or not. The following Theorem can thus be stated~:
\begin{theorem}
A general $N$-partite mixed state $\rho$ of rank 2 is separable if and only if all its $2 \times 2$ preconcurrence matrices $\tau_\alpha$, $\forall \alpha$, have their two singular values identical and are proportional to each other, in which case the mixed state $\rho$ is ensured to admit a separable decomposition over only two separable states.
\end{theorem}
It must be noted that for rank-2 mixed states the PPT criterion also provides a necessary and sufficient separability condition and allows one to prove that rank-2 mixed separable states admit a decomposition over only two separable states~\cite{Chen2013}. The current simultaneous hollowisability criterion provides in addition the separable decomposition of these states (see Theorem~\ref{genTh}).

For 2 qubits, since the system is entirely characterized by a single concurrence $C_\alpha$ and a single preconcurrence matrix $\tau_\alpha$, the simultaneous hollowisability question just comes down to a single hollowisability question and this is why the vanishing of the single concurrence $C_\alpha$ as computed via Eq.~(\ref{Calpha}) is enough to fully characterize the separability of the mixed states~\cite{Wootters}. For higher dimensional systems or systems composed by more than 2 parties, the number of generalized concurrences inevitably increases and the simultaneous hollowisability question cannot be avoided anymore.

We illustrate our NSC of separability with the nontrivial rank-5 3-qubit mixed state (expressed in the computational basis $|000\rangle, |001\rangle, \ldots$)
\begin{equation}
\label{rank5rho}
\rho = \frac{1}{20}\left( \begin{array}{cccccccc}
1 & -1 & 0 & 0 & -1 & 1 & 0 & 0 \\
-1 & 3 & 0 & 0 & 1 & -3 & 0 & 0 \\
0 & 0 & 6 & 0 & 0 & 0 & -2 & 0 \\
0 & 0 & 0 & 0 & 0 & 0 & 0 & 0 \\
-1 & 1 & 0 & 0 & 3 & 1 & 0 & 0 \\
1 & -3 & 0 & 0 & 1 & 5 & 0 & 0 \\
0 & 0 & -2 & 0 & 0 & 0 & 2 & 0 \\
0 & 0 & 0 & 0 & 0 & 0 & 0 & 0
\end{array}\right).
\end{equation}
For 3-qubit systems, we get the 9 independent operators
$S_\alpha = |000\rangle \langle 011| - |010\rangle \langle 001| + h.c.$,
$|100\rangle \langle 111| - |110\rangle \langle 101| + h.c.$,
$|000\rangle \langle 101| - |100\rangle \langle 001| + h.c.$,
$|010\rangle \langle 111| - |110\rangle \langle 011| + h.c.$,
$|000\rangle \langle 110| - |100\rangle \langle 010| + h.c.$,
$|001\rangle \langle 111| - |101\rangle \langle 011| + h.c.$,
$|000\rangle \langle 111| - |100\rangle \langle 011| + h.c.$,
$|000\rangle \langle 111| - |010\rangle \langle 101| + h.c.$,
$|000\rangle \langle 111| - |110\rangle \langle 001| + h.c.$
The resulting nine $5 \times 5$ preconcurrence matrices $\tau_{\alpha}$ are simultaneously hollowisable with the unitary matrix
\begin{equation}
U = \frac{1}{\sqrt{6}}\left( \begin{array}{ccccc}
-\sqrt{3} & 0 & 0 & 0 & \sqrt{3} \\
\sqrt{2} & 0 & -\sqrt{2} & 0 & \sqrt{2} \\
1 & 0 & 2 & 0 & 1 \\
0 & -\sqrt{3} & 0 & -\sqrt{3} & 0 \\
0 & -\sqrt{3} & 0 & \sqrt{3} & 0
\end{array}\right),
\end{equation}
out of which Theorem \ref{genTh} yields the separable decomposition $\rho = (1/5)\sum_{i=1}^5 |\psi_i\rangle\langle\psi_i|$,
with $|\psi_1\rangle = |\mathrm{-}01\rangle$, $|\psi_2\rangle = |\mathrm{-}0\mathrm{-}\rangle$, $|\psi_3\rangle = |10\mathrm{+}\rangle$, $|\psi_4\rangle = |\mathrm{-}10\rangle$, $|\psi_5\rangle = |010\rangle$,
where $|\pm\rangle \equiv (|0\rangle \pm |1\rangle)/\sqrt{2}$.
\section{Conclusion}
\label{Concl}

In conclusion, in this paper we first refined for general multipartite systems the necessary and sufficient condition of separability for pure states of Refs.~\cite{Yu2006,Audenaert} based on generalized concurrences. We showed how to obtain a minimal set of generalized concurrences to decide about the separability of an arbitrary multipartite pure state. We then showed that the general separability problem of mixed states is equivalent to a pure matrix analysis problem that consists in \emph{determining whether a given set of symmetric matrices is simultaneously unitarily congruent to hollow matrices, i.e., to matrices whose main diagonal is only composed of zeroes}. This mathematical reformulation of the quantum separability problem should pave the way towards new research in this field.

\acknowledgments
A.N. acknowledges a FRIA grant and the Belgian F.R.S.-FNRS for financial support. T.B. acknowledges financial support from the Belgian F.R.S.-FNRS through IISN Grant No. 4.4512.08.


\section*{References}

\begin{thebibliography}{10}

\bibitem{HorodeckiReview}
R.~Horodecki, P.~Horodecki, Horodecki M., and K.~Horodecki.
\newblock {\em Rev. Mod. Phys.}, $\textbf{81}$, 865 (2009).

\bibitem{Eck91}
A.~K. Ekert.
\newblock {\em Phys. Rev. Lett.}, $\textbf{67}$, 661 (1991).

\bibitem{Zei07}
R.~Ursin~\textit{et al.}
\newblock {\em Nat. Phys.}, $\textbf{3}$, 481 (2007).

\bibitem{Tei01}
A.~F. Abouraddy, B.~E.~A. Saleh, A.~V. Sergienko, and M.~C. Teich.
\newblock {\em Phys. Rev. Lett.}, $\textbf{87}$, 123602 (2001).

\bibitem{Wein10}
W.~Wieczorek, R.~Krischek, N.~Kiesel, Ch. Schmid, and H.~Weinfurter.
\newblock {\em Proc. SPIE 7608, Quantum Sensing and Nanophotonic Devices VII},
  76080P (2010).

\bibitem{Yu2006}
C.-S. Yu and H.-S. Song.
\newblock {\em Phys. Rev. A}, $\textbf{73}$, 022325 (2006).

\bibitem{GuhneToth}
O.~G\"uhne and T\'oth.
\newblock {\em Phys. Rep.}, $\textbf{474}$, 1-75 (2009).

\bibitem{Peres}
A.~Peres.
\newblock {\em Phys. Rev. Lett.}, $\textbf{77}$, 1413 (1996).

\bibitem{HorodeckiRealignment}
M.~Horodecki, P.~Horodecki, and Horodecki R.
\newblock {\em Open Syst. Inf. Dyn.}, $\textbf{13}$, 103 (2006).

\bibitem{Wocjan}
P.~Wocjan and M.~Horodecki.
\newblock {\em Open Syst. Inf. Dyn.}, $\textbf{12}$, 331 (2005).

\bibitem{Seevinck}
M.~Seevinck and J.~Uffink.
\newblock {\em Phys. Rev. A}, $\textbf{78}$, 032101 (2008).

\bibitem{HorodeckiPPT}
M.~Horodecki, P.~Horodecki, and R.~Horodecki.
\newblock {\em Phys. Lett. A}, $\textbf{223}$, 1 (1996).

\bibitem{Terhal}
B.~M. Terhal.
\newblock {\em J. Theor. Comp. Science}, $\textbf{287}$, 313 (2002).

\bibitem{HorodeckiLewenstein}
P.~Horodecki, M.~Lewenstein, G.~Vidal, and I.~Cirac.
\newblock {\em Phys. Rev. A}, $\textbf{62}$, 032310 (2000).

\bibitem{Chen2013}
L.~Chen and D.~Djokovic.
\newblock {\em J. Phys. A: Math. Theor.}, $\textbf{46}$, 275304 (2013).

\bibitem{Wootters}
W.~K. Wootters.
\newblock {\em Phys. Rev. Lett.}, $\textbf{80}$, 2245 (1998).

\bibitem{Audenaert}
K.~Audenaert, F.~Verstraete, and B.~De Moor.
\newblock {\em Phys. Rev. A}, $\textbf{64}$, 052304 (2001).

\bibitem{Badziag}
P.~Badzi\c{a}g, P.~Deuar, M.~Horodecki, P.~Horodecki, and R.~Horodecki.
\newblock {\em J. Mod. Opt.}, $\textbf{49}$, 1289 (2002).

\bibitem{Gillet}
J.~Gillet.
\newblock {\em Contribution to Entanglement Theory, Applications in Atomic
  Systems and Cavity QED}.
\newblock PhD thesis, University of Liege, 2011.

\bibitem{tensor}
T.G. Kolda and B.W. Bader.
\newblock {\em SIAM Rev.}, $\textbf{51}$, 455-500 (2009).

\bibitem{footnote}
We recall that the $k$-excitation Dicke states ($k = 0, \ldots, N$) are defined as
$|D_N^{(k)}\rangle = {N \choose k}^{-1/2} \sum_\pi |0, \dots, 0, 1, \dots, 1\rangle$,
where the multiqubit states in the sum contain $k$ qubits in state $|1\rangle$, and $\pi$ denotes all permutations of the qubits leading to different terms in the sum.

\bibitem{Hughston}
L.~P. Hughston, R.~Jozsa, and W.~K. Wootters.
\newblock {\em Phys. Lett. A}, $\textbf{183}$, 14 (1993).

\bibitem{Thompson}
R.~C. Thompson.
\newblock {\em Linear Algebr. Appl.}, $\textbf{26}$, 65 (1979).

\bibitem{Uhlmann}
A.~Uhlmann.
\newblock {\em Open Syst. Inf. Dyn.}, $\textbf{5}$, 209-227 (1998).

\bibitem{AlpinIkramov}
Yu.~A. Al'pin and Kh.~D. Ikramov, {\em Dokl. Akad. Nauk}, $\textbf{437}$, 7--8 (2011);
transl. in {\em Dokl. Math.}, \textbf{83} 141–142 (2011).

\bibitem{Gerasimova}
T.~G. Gerasimova, R.~A. Horn, and V.~V. Sergeichuk, {\em Linear Algebr. Appl.},
  $\textbf{438}$, 3829--3835 (2013).


\end{thebibliography}
\end{document}